\documentstyle[epsfig,12pt]{article}
\begin{document}

\begin{center}
\Large\bf {Physics of the Solar Cycle : New Views}
\end{center}
\begin{center}
{\Large\bf K. M. Hiremath}
\end{center}
\begin{center}
{\em Indian Institute of Astrophysics, Bangalore-560034, India,
E-mail : hiremath@iiap.res.in}
\end{center}

\begin{abstract}
With the recent overwhelming evidences that solar cycle
and activity phenomena strongly influence the
earth's environment and climate (Hiremath 2009a and references there in), it is necessary
to understand physics of the solar cycle and activity
phenomena. Genesis of the solar cycle and
activity phenomena-one of the major unsolved problem in solar physics-
 remains elusive to the solar community.
Presently there are two schools of thoughts viz.,
{\em turbulent dynamo} and {\em MHD oscillation} mechanisms
that explain the solar cycle and activity phenomena.
Both the mechanisms are critically examined and
fundamental difficulties are presented.
By keeping in mind the more advantages of having
MHD oscillation mechanism, compared to the
turbulent dynamo mechanism, following new ideas on
the genesis of the solar cycle and activity phenomena are presented.
Inevitability of most likely existence of a combined steady poloidal
and toroidal magnetic field structure in the solar interior. 
 Owing to suitable steady poloidal field structure, Alfven wave perturbations of 
long periods ($\sim$ 22 yrs) that excite in the solar core travel first to the 
poles in both the hemispheres and later reach the equator.
While traveling towards the surface, Alfven wave perturbations
along the weak poloidal field structure in turn
perturb the embedded strong toroidal field structure producing sunspots,
especially in the convective envelope, that travel to the surface
due to buoyancy along isorotational contours. With realistic density structure of the solar interior, 
computation of Alfven wave travel times along
different field lines of the poloidal field structure (Hiremath and Gokhale 1995)
yields almost similar periods ($\sim$ 22 yr) explaining the
constancy of 22 yr period of the odd degree modes obtained 
from the Spherical Harmonic Fourier analysis of the surface magnetic field. 
The observed quasi-periodicities of solar activity indices in 
the range of 1-5 years are explained due
to perturbation of the strong toroidal field
structure and, variation of very long period solar cycle and activity phenomena
such as the Maunder and grand minima is explained to be due to 
coupling of long period poloidal and toroidal MHD oscillations. 
\end{abstract}

\section {Introduction} 

Since the discovery of sunspots by Galileo, the physics of
solar cycle and activity phenomena is not understood
completely. There are two schools of thoughts-{\it turbulent dynamo
and MHD oscillatory mechanisms}-on the
genesis of the solar cycle and activity phenomena. Although turbulent dynamo models
explain qualitatively many of the observed
solar cycle and activity phenomena, there are several difficulties
and limitations in their application to the solar cycle
(Piddington 1971; 1972; 1973; Cowling 1981; Levy 1992;
Vainstein and Cattaneao 1992; Hiremath and Gokhale 1995 and references
there in; Hiremath 2001 and references there in). In this
talk, I revisit MHD oscillatory theory and show with
new ideas that many of the observed solar cycle and activity phenomena
can be explained. 

In section 2, I briefly summarize the important observations
related to solar cycle and activity phenomena. In section 3, both the turbulent dynamo 
mechanism and MHD oscillatory theory of the solar cycle are 
 critically examined. Salient features of Alfven theory on the solar
cycle are presented in section 4. In section 5, 
new views on the genesis of the solar cycle and activity phenomena
are proposed and important observations of solar cycle and activity
phenomena are explained. The conclusions of this study
are given in section 6.

\section {Summary of the Observations}

\subsection {Solar 11 Year Cycle and Grand Minima}

Variation of occurrence of the sunspots over the surface
of the sun with an average periodicity of $\sim$ 11 years is termed
as "{\em sunspot cycle}". The length of sunspot cycle also varies
between 9 to 12.5 years (Zwan 1981; Hiremath 2008a). Although sunspot activity
appears to be fairly regular (Dicke 1978; Hiremath 2006 and references there in), during
the period from 1645 to 1715, there was the dearth of 
sunspots and is called the Maunder minimum type
of solar activity. Sun might have witnessed such grand minima of solar 
activity during it's previous evolutionary history. 

\subsection {Solar 22 Year Magnetic Cycle}

Soon after the discovery of strong magnetic fields in sunspots,
Hale (1908) discovered that majority of the leading bipolar spots
in the northern hemisphere have the same polarity, whereas in the
southern hemisphere they are of the opposite polarity. 
These opposite polarities in both the hemispheres
will reverse during the next cycle.  Thus the period of
the solar magnetic activity cycle is twice as that
of a sunspot cycle and this phenomenon is called
"{\em 22 year solar magnetic cycle}".

\subsection {Sunspot Butterfly Diagram}

The sunspots' occurrence in a particular latitude
belt varies between nearly $40^{o}$ north-$40^{o}$
south of the equator. During the beginning of
a solar cycle, sunspots of a new cycle appear at
the higher latitudes. As the cycle progresses,
the occurrence of sunspots drift towards the solar
equator from the higher latitudes of both the
northern and southern hemispheres constituting
so called the {\em sunspot butterfly diagram}.

\subsection {Spherical Harmonic Fourier Analysis of Magnetic Activity}

Spherical harmonic Fourier (SHF) analysis (Stenflo and Vogel 1986;
Stenflo 1988; Knack and Stenflo 2005) of magnetograms
and inferred magnetic field (Gokhale, Javaraiah and Hiremath 1990;
Gokhale and Javaraiah 1992) from the sunspots data show
 that the axisymmetric global oscillations with specific
periods ($\sim$ 22 yr and smaller) do contribute
predominantly to the evolution of the large-scale
photospheric magnetic field. The power spectra
of these data show that the {\em odd}
and {\em even} parity modes behave differently. All the
odd parity modes have same periodicity of $\sim$ 
22 years and the frequency of even parity modes
increases with degree $l$ that is almost similar
to the observed helioseismic $p$ mode spectrum.

\section {Theoretical Models of the Solar Cycle}

In case of the sun, the diffusion time scale of large-scale
global magnetic field structure is $\sim$ few billion years,
i.e., greater than the sun's age itself. Hence the sun
is expected to retain some of it's primordial
magnetic field structure (Hiremath and Gokhale 1995 and references
there in) and varies on time scale much larger
than the dynamical time scale. Hence it is easy
to understand the existence of observed magnetic field
structure, if it were found to be steady with time.
However, the large-scale field observed at the surface 
varies in a cyclic manner with time scales of $\sim$ decades.
Thus one has to seek the theoretical frame work 
that not only incorporates the maintenance of
magnetic field structure but also it's periodic
behavior. That means one needs some sort
of a dynamo mechanism-that maintains electro magnetic
field against dissipation at the cost of energy 
provided by some source-in the interior of the sun.

Presently there are two main schools of thoughts
on the modeling of solar cycle : (i) Turbulent dynamo
mechanism and, (ii) MHD oscillatory theories.

\subsection {Theory of Dynamo Mechanism}

These theories are based on the fact that moving conductors
generate electric currents due to electromagnetic induction.
It is therefore expected that in the sun the flows like
rotation and convection could provide dynamo action through
electromagnetic induction. However, all the velocity
fields can not maintain the dynamo. For example, according
to Cowling's (1934) theorem, steady axisymmetric magnetic
fields can not be maintained by axisymmetric flows. Thus
the dynamos for sun-like stars ought to be either
non-axisymmetric or non-stationary (or both).
 
In this mechanism, the dynamo effect is statistically
averaged over the turbulent flows. The velocity ${\bf u}$ 
and the magnetic field ${\bf B}$ of the plasma are expressed
as sums of mean part ($<u>$ and $<B>$-large-scale 
and slowly varying terms) and the fluctuating parts ($u'$
and $B'$-small scale and rapidly varying terms).
With certain assumptions and approximations, the final
equation that governs the spatio-temporal variation of the
average magnetic field is of the form 
\begin{equation}
{{\partial {\bf <B>}}\over{\partial t}} =
\bf{curl} [{\bf {\alpha <B>}} +{\bf <u>} x {\bf <B>}] - curl[({\bf \eta + \beta})curl<B>] \,
\end{equation}

\noindent where $\alpha$ is helicity and $\beta$ is diffusivity 
due to turbulence and $\eta$ is electromagnetic diffusion.

Further, the mean field components ($<B>$ and $<u>$)are written in terms of
the poloidal (magnetic field $B_{p}$ and 
velocity, i.e., meridional flow) and the toroidal 
 (magnetic field $B_{\phi}$ and the angular velocity i.e., $\Omega$ ) components leading to two 
equations containing $\alpha$, $\beta$ and $\eta$ as the free parameters.
If one knows the internal rotation $\Omega$ and the
meridional velocity, with specifying free parameters,
in principle, one can evolve two equations and 
one can reproduce the solar butterfly diagrams (Hiremath
and Lovely 2007 and references there in). These dynamo
models are called ``{\em kinematic dynamo models}". 

Initially, a mechanism for the production of sunspots was
proposed by Cowling (1953), who suggested that sunspots 
are eruptions of submerged toroidal fields produced by
the differential rotation acting on a weak poloidal
field. Subsequently, Parker (1955) proposed that the 
poloidal field itself is regenerated by the interaction
between cyclonic convection and buoyantly rising toroidal flux
elements. Incorporating these ideas, Babcock (1961)
phenomenologically modeled the solar cycle. Leighton (1964, 1969)
presented a semi-empirical model of the solar
cycle and reproduced the well known sunspot butterfly diagram.  
Recently we have excellent  reviews (Charbonneau 2005; Dikpati 2005;
Brandenburg and Subramaniyam 2005; Solanki, Inhester and Schussler 2006;
Hiremath and Lovely 2007 and
references there in; Choudhuri 2008) on the solar dynamo mechanism. 

\subsubsection {Difficulties in the Turbulent Dynamo Models
of the Solar Cycle }

Though the turbulent dynamo models of the
solar cycle reproduced elegantly the properties of
the solar cycle and activity phenomena, the models plague with many
fundamental difficulties (Hiremath 1994 and references there in;
Hiremath 2001 and references there in; Petrovay 2000; Hasan 2008;
Venkatakrishnan and Gosain 2008; Choudhuri 2008; Nandy 2009) and few of them are presented
in the following.
\begin{itemize}

\item The ``{\em first order smoothening  approximation}"
used for the derivation of the induction equation
is valid only when the fluctuating field is very much 
smaller than the mean field. This is possible only when
(a) the eddy magnetic Reynolds number $R_{m}$ is $<<$1,
and (b) the correlation time $\tau$, the eddy of length $\lambda$,
and the r.m.s velocity $v$ are related as $\tau<<{\lambda/v}$.
In reality, neither of these conditions is valid on the sun
where $R_{m}>>1$ and $\tau \sim {\lambda/v}$.

\item  According to Piddington (1971; 1972; 1973) : (a) the concept
of turbulence of the solar magnetic field is unsound; turbulence
may mix magnetic elements but does not destroy large-scale
magnetic fields. In fact recent MHD simulations (Brun {\em et. al.} 2004;
Stein and Nordlund 2006; Jouve and Brun 2007; 
Bushby et a.l. 2008; Steiner et. al. 2008; Miesch and Toomre 2009 and
references there in) 
 of the convective envelope substantiate the
Piddington's ideas; (b) the field created by the eddy motions
would be mainly turbulent field, unlike the field that is
actually observed; (c) the field created during successive
cycles would rise successively to higher levels and the
whole field would eventually leave the sun.

\item  Except the previous study of Brandenburg (1988) that possibly
explains the diagnostic power spectrum of the even degree
modes of the SHF analysis (Stenflo and Vogel 1986) of magnetic field, so far,
no dynamo mechanism explains the constant power of 22 years
in the odd degree modes whose superposition represents the large-scale 
solar cycle and activity phenomena (Stenflo 1988; Stenflo and Gudel 1988;
Gokhale and Javaraiah 1992).

\item The values of parameters $\alpha$, $\beta$ and of
the rotational shear are either arbitrarily chosen or
estimated crudely from the statistical properties 
of the observed motions. In order to reproduce sunspot butterfly diagrams, 
dynamo models require substantial increase of rotational profile
from surface to the interior contradicting the rotational
profile as inferred by the helioseismology (Hiremath 1994; Hasan 2008).

\item Within the framework of kinematic dynamo models, it is 
impossible to address the question of limiting amplitude
of generated magnetic flux owing to linearity of the
induction equation. 

\item One of the fundamental problem in keeping the solar
dynamo in the convection zone is buoyant rise of all
the flux on time scales very much smaller than the
solar cycle period. This difficulty can be avoided
if the dynamo process is operating in a stably
stratified region beneath the solar convection
zone. However, as pointed out by the previous
studies (De Luca and Gilman 1991; Hasan 2008), 
the process of dynamo mechanism operating
beneath the solar convection zone could add some
other serious difficulties. For example, how the
magnetic flux injected into the convection zone
is a question. 

\item In order to reproduce proper solar butterfly diagrams 
(Hathaway et. al. 2004) and predict future solar cycles, 
the flux transport dynamo models (Nandy and Choudhuri 2002;
Dikpati and Gilman 2006; Choudhuri, Chatterjee and Jiang 2007) 
require the meridional circulation that needs to penetrate 
 below base of the convective envelope. However, recent
studies (Gilman and Miesch 2004; Svanda, Kosovichev and Zhao 2007;
Hiremath 2008b) conclude that owing to high density
stratification and strong ($\sim$ $10^{4}$ G) toroidal magnetic field structure,
it is very unlikely that the return flow will reach the surface
 with a period of solar cycle. 
\end {itemize} 

\subsection { Theory of MHD Oscillations}

In the electrically conducting magnetized plasma,
there are three kinds of MHD (magnetohydrodynamic) waves,
viz., (i) Alfven wave, (ii) slow MHD wave
and, (iii) fast MHD wave.

Since the sun is such a dynamic body that always disturbances in the medium
exist.  Such disturbances perturb the magnetic field structure
leading to generation of Alfven waves.
 Alfven waves are of two types (Priest 1981), {\em viz.,}
{\em shear Alfven waves} due to incompressibility
 and {\em compressible Alfven waves} due to compressibility.
The {\em shear Alfven waves} are transverse
waves that travel along the field lines, where as the
{\em compressible Alfven waves} consist of both longitudinal
and transverse waves. Since the time scales of compressible
waves ($\sim$ 5 min) due to density perturbations are very much smaller than
the solar cycle time scales ($\sim$ 22 yrs), the
condition of incompressibility applies and {\em shear Alfven waves}
are best suited for the present study.

Observed periodic behavior of the large-scale magnetic field
structure of the sun is viewed as a consequence of MHD (magnetohydrodynamic)
oscillations in the presence of a large-scale steady (diffusion
time scale  $\sim$
billion years) magnetic field structure. These theories recognize
the fact that most of the observed fields at the surface (including
 those in the polar regions), are in the form of bipolar
regions. The MHD oscillations must be azimuthal perturbations
of ambient steady poloidal magnetic field structure. The amplification of 
the toroidal field can results from the azimuthal perturbations
of the ambient steady poloidal magnetic field.
Any such perturbations of the field lines would eventually lead
to MHD waves. The waves travel along the field lines of the steady poloidal field
structure and are reflected due to density gradients
near the surface. Superposition of these traveling waves
lead to stationary or standing oscillations. The strong
fields needed for activity result from the constructive
interference of these waves.

For an axisymmetric magnetic field structure and in cylindrical
geometry, the MHD wave equation (Mestel and Weiss 1987) is 
given by

\begin{equation}
{\partial^{2} { \Omega }\over{\partial t^{2}}} = {{B_{p}}^2\over{4\pi\rho}}{\partial^{2} {\Omega }\over{\partial s^{2}}}
\end{equation} 
where ${\Omega}$ is angular velocity, ${B_{p}}$ is poloidal
component of the steady magnetic field structure and 
$\rho$ is the density of the ambient plasma. 
In addition we have a similar equation
by replacing ${\Omega}$ by ${B_{\phi}}$.
These two equations imply
that the changes in either $\Omega$ or
 $B_{\phi}$ propagate with the local Alfven speed
$V_{A}={B_{p}}/\sqrt{ {4\pi\rho}}$ determined by the steady 
poloidal field structure. Since the perturbations
are in the azimuthal direction, such a wave equation
is called {\em torsional MHD wave equation}.
 In fact, in the following subsection, we
use this equation for checking the admissibility of
global torsional MHD oscillations in various models
of the steady magnetic field structures in the solar interior. 

Alfven (1943) and Walen (1949) were the pioneers to
propose this theory and latter their ideas were revived
by many authors (Layzer et. al., 1955; Plumpton and
Ferraro 1955; Piddington 1976; Layzer et. al., 1979,
Vandakurov 1990; Hiremath 1994; Hiremath and Gokhale
1995). In the following, first we revisit Alfven's seminal work on
the theory of solar cycle and also present the fundamental
difficulties.

\section { Alfven's Theory of Solar Cycle}
In his seminal work, Alfven (1943) assumed that : (i)
the sun consists of large-scale dipole magnetic field
structure in the interior whose magnetic axis coincides
with the rotation axis,
(ii) a magnetic disturbance somewhere else in the
deep interior travels with Alfven speed $V_{A}$ along
the field lines and reach the surface,
(iii) excitation of MHD waves is due
to turbulence that is created by the differences in
the velocity gradients of the isorotational contours and,
(iv) coupling between neighboring field lines expected
to transfer the oscillations towards all parts of
the sun, 

For a polytropic density variation, and for the dipole
magnetic field structure with a dipole moment
$\sim$ $4.2X10^{33}$ G $cm^{3}$, Alfven computed the travel
times along different field lines and found that $\sim$ 70
years for the field lines near the pole, and  $\sim$ 80
years for the field lines near the equator. Since these
periods did not agree with the 22 year period, he concluded
that the 22 year period must be the resonance period
of some lines of force in the interior. In addition,
Alfven's theory also explained the observed propagation
of sunspot zones and opposite polarities of the sunspots.

 Alfven computed dependence of the sunspot frequency
 with respect to latitude and found almost similar
results as that of observation. By the theory
of standing oscillations along different field lines, 
Alfven explained the observed fact that during a
particular cycle the sunspots
in both the hemispheres have opposite polarities. Assuming
that the perturbations in the interior are irregular, 
he made an attempt to explain the long period
sunspot activity.
\subsection {Difficulties in the Alfven's theory of
Solar cycle}

Though Alfven's theory appears to explain
most of the observations of the solar cycle and activity
phenomena, following are important difficulties :
(i) assumed polytropic density stratification and 
magnetic dipole field structure that has a
singularity near the center are unphysical, (ii)
computed period for the one hemisphere is $\sim$ 40 years, nearly four
times the period of the sunspot cycle; independent of Alfven's work recently
 Davila and Chitre (1996) also computed the travel 
times for the assumed radial field and obtained
the travel time of $\sim$ 300 years for the fundamental
mode, (iii) although origin of sunspots
is proposed to be due to superposition of 
long period oscillations, more observational
and theoretical inferences are needed, (iv) intensity
of the sun's apparent dipole field is assumed to be
$\sim$ $10^{3}$ G contradicting the observations ($\sim$
 1 G), (v) it is not clear how the random perturbations
lead to dearth of sunspot activity, similar to 
Maunder minimum, (vi) if one accepts the Alfven's 
model of magnetic field structure in the solar interior, 
it is not possible to reproduce important result
(odd degree parity modes have constant period of 
22 years) from the SHF analysis of the observed surface
magnetic field (Stenflo and Vogel 1986; Knack and Stenflo 2005)
and the field inferred from the sunspots (Gokhale, Javaraiah
and Hiremath 1990; Gokhale and Javaraiah 1992).
{\em All the afore mentioned difficulties of Alfven's
theory suggest a suitable geometrical magnetic field
structure with proper intensity in the solar interior.}

\section {New Ideas on physics of the Solar Cycle}
Firstly, we have to admit that MHD oscillatory theories 
have the following three main
difficulties : (i) the lack of observational evidence of magnetic
field structure
of primordial origin, (ii) difficulty in believing that such a perturbed poloidal field
structure of weak general magnetic field ($\sim 1\, G$) can produce sunspot
activity of strong magnetic field ($\sim \, 10^{3}\, G$),
 (iii) owing to strong dissipation in the convective envelope, long
period ($\sim$ 22 years) MHD oscillations can not be maintained for the next cycle.

\subsection {Existence of a combined Poloidal and Toroidal Magnetic Field Structure in the solar
interior}

Likely existence of large-scale poloidal magnetic
field structure can be confirmed from the white light pictures 
(see the Fig 1 of Ambroz {\em et.al.} 2009; Pasachoff 2009; 
see the Fig 3 and 8 of Pasachoff {\em et. al.} 2009) during 
total solar eclipse around solar minimum.
Though direct observational measurements of such a large-scale
 weak magnetic field ($\sim$ 1 G) are lacking, indirectly, 
from the helioseismic rotational
isocontours we (Hiremath 1994; Hiremath and Gokhale 1995) proposed
 a most likely poloidal magnetic field structure of primordial
origin in the solar interior.

Observations show that strength of the poloidal field
is very weak ($\sim$ 1 G) compared to the strength of 
rotation, hence the poloidal field must isorotates with the
internal rotation of the plasma. This implies that geometrical
poloidal field structure must be similar
to the geometrical structure of the internal isorotational contours as inferred 
from the helioseismology. In fact it is true for
the rotational isocontours (as inferred from the helioseismology)
in the convective envelope where inferred rotational
isocontours are reliable.  In previous study and by using 
Chandrasekhar's MHD
equations, we (Hiremath 1994; Hiremath and Gokhale 1995)
modeled the steady part of the poloidal field structure
and found the diffusion time scale to be $\sim$ billion years.
Gough and McIntyre (1998) also have proposed the inevitability 
of such a poloidal field structure in the radiative interior.
With reasonable assumptions and approximations and,
by using MHD equations, we (Hiremath 2001)
consistently obtained solution for both internal rotation
and toroidal component of the magnetic field structure
in the convective envelope.
The toroidal field structure in the convective
envelope has a quadrupole filed like geometric
structure and the field strength varies from $\sim$
 $10^{4}$ G near base of the convection zone to
$\sim$ 1 G near the surface. 
For the sake of stability (Mestel and Weiss 1987; Spruit 1990;
Braithwaite and Spruit 2004), such a combined poloidal and toroidal
field structure is necessary in the solar interior.

Hence, the sun may be pervaded by the combination of large-scale
steady poloidal and toroidal magnetic field structures
(both of which may of primordial origin and diffusion
time scales are $\sim$ billion years). If one accepts the  
existence of such a combined field structure, the first
difficulty in MHD oscillatory theory can be removed.

\subsection {Genesis of the Solar Cycle and Activity Phenomena}

The second difficulty of the oscillatory model can be removed as follows.
Fallowing Alfven (1943) , any perturbations near the center travel along and
perpendicular to the poloidal field structure and,
coupling between neighboring field lines transfer
the perturbed energy to all parts of the sun. 
 The interesting property of the {\em shear Alfven
waves} is that the magnetic and velocity perturbations
are perpendicular to the magnetic field lines and travel
along the field lines. That means the Alfven waves while traveling
along the field lines perturb in turn the neighboring field lines.
If one believes that the sun has a magnetic field structure
similar to one proposed by Hiremath and Gokhale (1995),
then the field lines that pass through north and south poles (the
field line represented by `A' in Fig 1 of Hiremath and Gokhale (1995))
in both the hemispheres experience the Alfven wave perturbations first and the field lines
that are close to the equator (the field line represented
by `L' in Fig 1 of Hiremath and Gokhale (1995)) experience the Alfven wave perturbations
later. Thus there is a phase lag of $\pi/2$ radians between the
polar and equatorial solar activities. This reasoning that
Alfven wave perturbations reach first poles and then equator is 
consistent with analysis of the sunspot butterfly diagrams 
(Pelt et al 2000), the observations of torsional oscillations
on the surface (Howard and La Bonte 1980; Komm, Howard and Harvey (1993)), 
theoretical (Hiremath 1994) and helioseismic inferences
(Zhao and Kosovichev (2004); Antia, Basu and Chitre (2008) and, 
in the atmosphere (Altrock, Howe and Ulrich 2008).

Perturbations of the poloidal field structure in the convective
envelope in turn perturbs the embedded toroidal
field structure and, superposition of many such
azimuthal perturbations attains a critical strength leading to
formation of the sunspots and due to buoyancy raise
along the isorotational contours and reach the 
surface. 
 For example, if one accepts the existence of such a steady part
of toroidal magnetic field structure with a
strength $B_{\phi}$, then perturbations result in creation of MHD
waves whose amplitudes are $\sim$ $\delta B_{\phi}$. Superposition
of many such MHD waves in turn leads to constructive
interference and form the sunspots and, erupt towards the surface
along the isorotational contours. As for the reversal of polarity,
 once sunspots are formed, they raise towards the surface
in a particular latitude belt due to buoyancy and meridional flow
transports the remnant of the flux on the surface
towards the poles and change the sign.

Due to turbulence in the convective
envelope, the amplitude of the Alfven wave perturbations that travel along
the poloidal field (isorotational contours) will be considerably
reduced near the surface. That means there is a need of
constant forcing for every 22 years near the center.
Hence, it is not surprising that the resulting 11 year
 {\em solar cycle and activity phenomena
on the surface can be
considered as a forced and damped harmonic oscillator} (Hiremath 2006).
In this way the third difficulty of theory
of MHD oscillations can be removed.

\subsubsection{Implications for the combined poloidal and toroidal field structure }

 Some other consequences of having such a steady toroidal magnetic field 
structure in the convective envelope are :
(a) perturbations to the thermal sound speed
in the solar interior that contributes to splitting
of the even degree $p$ modes (Basu 1997; Antia, Chitre and Thompson 2000;
Antia 2002); (b) explanation for the recent discovery of ubiquitous horizontal magnetic field
structure in the quiet-sun internetwork regions pervading every 
where in the photosphere as detected by Hinode
satellite (Jin, Wang and Zhou 2009; Wijn  {\em et. al.} 2009 and references
there in; Tsuneta {\em et. al.} 2009; Lites {\em et. al.} 2008; 
Lites {\em et. al.} 2009) and ground based telescope ( Lites {\em et. al.}  1996;
Beck and Rezaei 2009); (c) Alfvenic perturbations of the poloidal field
structure (Hiremath 1994; Hiremath and Gokhale 1995) should yield
the periods around 22 years and of the toroidal
field structure should yield the periods around 1-5 years.

\subsubsection{Physics of the 1-5 year quasi periodicities }

 As for the steady toroidal field structure, the periods are 
computed from the relation $ \tau \sim {L \over V_{A}}$,
where $\tau$ is the period of oscillations, $L$ is length scale of the field
lines and $V_{A}$ is the Alfven velocity. In case of the toroidal
field structure, the length $L$ is considered to be $\sim$ 2$\pi \,r$,
where $r$ is the radius of the ring along the azimuthal
direction. For example, at radius of 0.1$R_{\odot}$, perturbation of the ring of 
toroidal field structure with a intensity
$10^{5} G$ (and  density of $\sim$ 150 $gm/cm^{3}$) yields the period of $\sim$ 5
years. If we accept the model (Hiremath 2001) of steady part of
toroidal field structure (with a intensity $\sim$ $10^{4}$ G near
base of the convection zone and $\sim$ 1 G near the surface) 
in the convective envelope and by taking
the typical density values, the period of the oscillations 
vary from $\sim$ 1.3 years near base of the convective 
envelope to $\sim$ of few months near the surface. 
These physical inferences imply that as the Alfven wave
perturbations travel along different field lines ( or along different isorotaional
contours) of poloidal field structure and reach the surface from pole to equator, one would expect 
periodic phenomena at a particular latitude zone on the surface that is 
connected with periodic phenomena at a particular radius in the solar interior.
To elaborate further, from the above inferences and with the poloidal field
structure (between the field lines zone represented by the
symbols A-C of Fig 1 of Hiremath and Gokhale 1995), one would expect near 5 year periodic
phenomena, that originate in the beginning of solar
cycle and at radius of 0.1$R_{\odot}$, should occur
at the higher latitude zones. Similarly near 1.3 yr periodic phenomena
that occur near base of convection zone ( between the field lines zone 
represented by the symbols I-J of Fig 1 of Hiremath and Gokhale 1995)
travel along the field lines and reach the surface around
solar cycle maximum and in the 20-25 deg latitude (or 70-75 deg
 colatitude) zone on the surface. To conclude of this subsection,
in addition to 11 yr periodicity in both the hemispheres,
near 5 and 1.3 yr periodicities should occur during certain
phase of the solar cycle. From the observed periodic analysis
of different solar activity indices, let us examine in the following
whether conclusion of this subsection is right or wrong.                                                                    

Observations show that near 5 and 1.3 yr periodicities
are indeed quasi-periodic and occur at different
epochs (or at different latitude zones on the surface)
of the solar cycle. For example near 5 quasi-periodicity is
detected in the high latitude zones (Vecchio and Carbone
2009 and references there in). Although near 11 yr periodicity is 
dominant in the analysis of high latitude filaments 
(Li {\em et. al.} 2006), near 5 yr periodicity has
a very low spectral power in their analysis.

As for near 1.3 yr periodicity, it is detected in the sunspot data (Krivova and Solanki 2002),
in the photospheric mean rotation (Javaraiah and Komm 1999; Javaraiah 2000),
in the magnetic fields inferred from H-alpha filaments (Obridko and Shelting 2007),
in the large-scale photospheric magnetic fields (Knaack, Stenflo and Berdyugina 2005),
in the green coronal emission line (Vechhio and Carbone 2009) and in
the occurrence of coronal mass ejections (CME) (Hiremath 2009b).
Spherical harmonic Fourier analysis (Stenflo and Vogel 1996;
Knaack and Stenflo 2005) of
magnetograms taken over 22 years shows the combined powers for the period of 22 years
(due to a weak poloidal field $\sim$ 1 G (Stenflo 1994)) and 1-5 years (due to a strong toroidal
field of strength $\sim\, 10^{4}-10^{5}$) respectively.
From the helioseismic data, 1.3 yr periodicity is detected near base of the
convection zone (Howe, et. al., 2000; Howe 2009). However, using same helioseismic data,
Antia and Basu (2000) conclude that there is no 1.3 yr periodicity near base
of the convection zone. Further analysis of  Basu and Antia (2001) shows
somewhat similar period as reported by Howe et al. (2000) but did not 
consider it to be significant. Interestingly, as expected by this study,
analysis of post-2001 data (Toomre et al., 2003; Howe et al., 2007;
, see Figure 32 of Howe 2009) shows the disappearance of 1.3 yr periodicity.
Some more such data analysis are required in order to confirm the 
physical inference of this study that 1.3 yr
quasi-periodicity (that occurs around solar maximum) is the result of periodic disturbances  
near base of the solar convective envelope.

\subsection {The Nature of Drivers}
In order to maintain the 22 yr oscillations for each
solar cycle, either internal or external driver that re-excite the
oscillations is necessary. When we say the driver, we mean
the unknown perturber that perturbs the magnetic field lines
 near the center periodically. We don't know the nature
and origin of the drivers.

When we say ``internal driver", we mean the driving mechanism near
the solar center due to local perturbations. Since the sun
is such a dynamic body that always disturbances in the medium
exist.  Such disturbances perturb the magnetic field structure
leading to generation of Alfven waves. One such disturbance
is the local thermonuclear runways as proposed by Grandpierre and G'bor (2005).

On the other hand, ``external driver" means driving
due to combined gravitational forces of the solar system
near the solar center or tidal forces due
to planets (Javaraiah and Gokhale 1995 and references
there in; Wilson, Carter and Waite 2008). 
According to Zaqarashvili (1997), sun's motion around the solar
system barycenter causes the weak periodic differential rotation
that shears the poloidal field periodically leading to 22 year Alfvenic
oscillations. However, there are studies (De Jager 2005; De Jager
and Versteegh 2005; Shirley 2006) that rule out the
external perturbation of driving solar cycle
and activity phenomena. Hence, at the present stage, it is very difficult to
delineate which driver drives the 22 yr oscillations.
However, detection of solar internal gravity (`g') modes
(Unno {\em et. al.} 1979; Hiremath 1994; Christensen-Dalsgaard 2002;
Christensen-Dalsgaard, J., 2003; Garcia {\em et. al.} 2008; Jimenez and Garcia 2009 
and references there in) will definitely delineate 
these unknown drivers that excite 22 year oscillations.

\subsection {Alfven wave Travel Times}

From the SHF analysis of the sun's magnetic field,
it is found that the axisymmetric terms of odd parity modes
have nearly the same periodicity ($\sim$ 22 years). This 
indicates that the Alfven wave travel times may be approximately
same along different field lines of a steady magnetic field
structure. In order to check the admissibility of such global
oscillations, we have computed the Alfven wave travel times
$\tau =(ds/V_{A})$ along different field lines (that
originate at the center and cut across the surface 
from pole to the equator), where $ds$ is the line element of the
magnetic field structure and $V_{A}$ is the Alfven wave
velocity. Alfven wave travel
times are computed in the following models by taking into account the
real density variation in the sun : (i) the uniform field,
(ii) the dipole field, (iii) the combination of uniform
and dipole field, (iv) the combination of dipole and 
hexapole embedded in a uniform field (Gokhale and Hiremath 1993)
and, (v) solution of a diffusion equation in an incompressible
medium of constant diffusivity (Hiremath and Gokhale 1995).

For the sake of comparison, all the models are assumed to have
the same amount of magnetic flux with a nominal value
of $1.5X10^{22}$ Mx corresponding to
a uniform field of $\sim$ 1 G. It is found that, for all the field lines, the last
two models yield the same period of 22 years. It is concluded that, owing
to regularity (without singularity) of the magnetic field structure at the center, 
the last model can be most likely 
the suitable geometrical magnetic field structure that
sustains near 22 years oscillations for all the field
lines explaining the constancy of $\sim$ 22 years
 of the observed SHF analysis of odd parity modes. 
 
\subsection {Coupling of Poloidal and Toroidal MHD oscillations
and the Maunder Minimum type of activity}

As mentioned in section 2.1, sun might have experienced the 
dearth of sunspot activity in the past evolutionary
history. Yet there is no complete consensus among
the solar community  whether such grand minima are 
chaotic or regular. However, in the previous study
(Hiremath 2006, end of section 3), it is concluded
that Maunder minimum type of activity is not chaotic and must
be periodic with a period of $\sim$ 100 years. Although
most of the dynamo mechanisms (Choudhuri 1992; Charbonneau 
and Dikpati 2000; Usoskin, Solanki and Kovaltsov 2007; Moss et. al. 2008;
Brandenburg and Spiegel 2008) treat the long term variations
of the solar cycle and activity phenomena as chaotic, based on the
previous studies (Feynman 1983; Price, Prichard and Hogenson 1992;
Hiremath 2006 and references there in), we consider 
such a long term solar cycle and activity phenomena to be periodic.
\begin{figure}[h]
\begin{center}
\psfig{figure=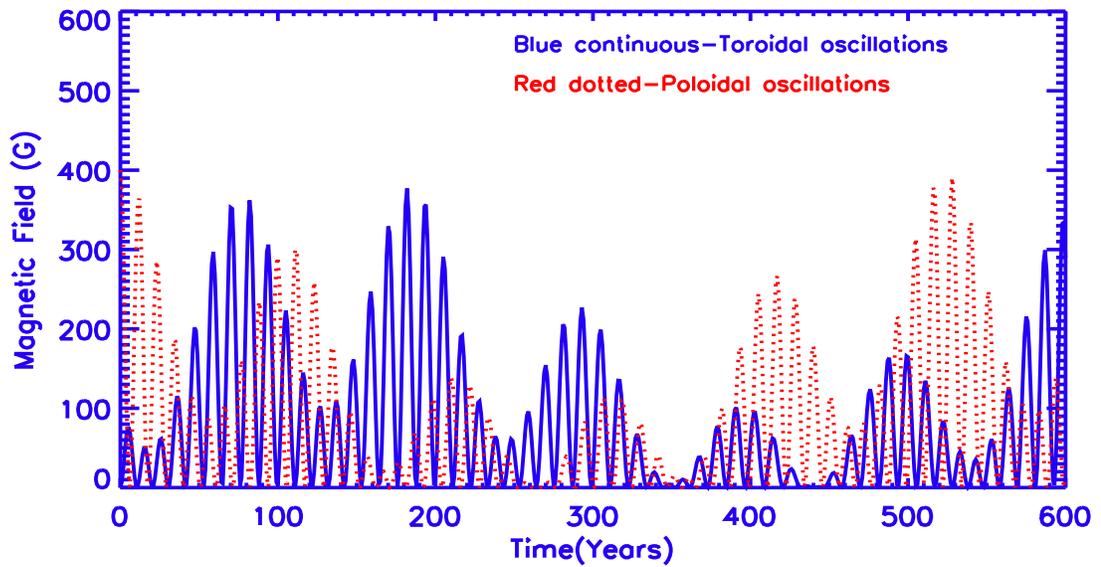,height=8.0cm,width=15.0cm}
\end{center}
\caption{
The sun's long period coupled oscillations of the poloidal 
and toroidal magnetic field structures.
 The sunspot activity that results from the superposition
of toroidal field oscillation modes is represented by blue continuous line and the
poloidal field oscillations is represented by the red dotted line.
}
\end{figure}

In the previous study, the observed solar cycle is modeled
as a forced and damped harmonic oscillator that consists
of sinusoidal and transient parts. It is found 
that the simultaneous change in magnitude
of phase difference ($\sim \, 2\pi$ radians) between the transient
and sinusoidal parts  and of very low sunspot
activity may be due to the Maunder minimum type of oscillations. 
This result possibly suggests the following : either a beat phenomenon
due to close frequencies or coupling of long period poloidal and toroidal
MHD oscillations. Although beat phenomenon can yield the
Maunder minimum type of lull of activity, the constant 
amplitude of the beat activity can not match varying
long term period amplitudes as shown by the observations. On the other hand,
as presented below, profile of coupled poloidal and toroidal 
oscillations is almost similar to the observed long-term variation
of the solar activity that constitutes Maunder and other
grand minima. Following Fletcher and Rossing (1998), on the theory
of mechanical vibrations, analytical solution of the equations governing 
the coupled oscillations 
of the poloidal ($B_{P}$) and toroidal ($B_{T}$) magnetic field structures 
in the dissipative medium  is derived as follows.

\begin{equation}
B_{P} = a_{0}cos(w_{0}t)+a_{1}cos(({w_{2}-w_{1} \over{2}})t)cos(({w_{2}+w_{1} \over{2}})t) \, 
\end{equation}
\begin{equation}
B_{T} = a_{0}cos((w_{0}+\pi/2)t)+a_{1}sin(({w_{2}-w_{1} \over{2}})t)sin(({w_{2}+w_{1} \over{2}})t) \, 
\end{equation}

\noindent where $t$ is time variable, $a_{0}$ and $a_{1}$ are
the amplitudes of the oscillation due to poloidal field and coupled
oscillations, $w_{0}={2\pi}/T$, T is period due to poloidal field, 
$w_{1}=w_{0}\sqrt{1-(\gamma/w_{0})^{2}}$,
$w_{2}=w_{0}\sqrt{1+2(w_{c}/w_{0})^{2}-(\gamma/w_{0})^{2}}$, 
 $w_{c}=2\pi (\sqrt{{V_{AP}}^{2} \pm  {V_{AT}}^{2}})/\delta R$ is coupling
frequency due to poloidal and toroidal oscillations,
$V_{AP}$ and $V_{AT}$ are Alfven wave velocities due to poloidal
and toroidal magnetic field structures and $\delta R$ is distance
between the neighboring field lines. 
The first term in the RHS of equations (3) and (4) is oscillation 
due to poloidal magnetic field structure and second terms in the RHS
of both the equations are coupling of oscillations due to
both poloidal and toroidal field
structures with a coupling frequency $w_{c}$. 

In order to closely match with the 11 year cycle and the long term 
variation of the sunspot activity,
the fundamental period due to poloidal oscillations must be 22 years
(or frequency $\omega_{0}$ is $\sim$ 0.286 rad/yr),
the dissipation factor $\gamma$  must be 0.185 and the coupling frequency
$\omega_{c}$ should be 0.11 rad/yr. It is interesting
to be noted that theoretical dissipation factor $\gamma$ of 0.185
is almost same as the dissipation factor of 0.186 obtained
from the observed solar cycles (Hiremath 2006). The simulation of 
magnetic energy (square of amplitude of either poloidal
or toroidal oscillations with arbitrary and equal amplitudes of $a_{0}$ and $a_{1}$)
of such coupled oscillations with respect to
 time span of 500 years (Fig 1) shows that oscillations
of the poloidal field with a fundamental period of 22 yrs
 excite the toroidal field
oscillations such that the toroidal field structure oscillates in consonance
with the poloidal field oscillations resulting in coupling of poloidal and toroidal
oscillations that reproduce the observed cyclic
periodicities of 11  and 100 yrs with
a very deep minimum around 350 years when both the strengths of poloidal
and toroidal oscillations have very low amplitudes. 
The paleoclimatic records show that during Maunder minimum although the sunspot
activity was practically absent, the 11 year activity
due to geomagnetic indices (Cliver, Boriakoff and Bounar 1998) 
and solar proxy records (Beer, Tobias and Weiss 1998; De Jager 2005;
Muscheler {\it et al} 2007) was
present. As the activity of geomagnetic indices (Feyman 1982; 
Legrand and Simon 1991; Georgieva and Kirov 2006)
and the solar proxy records are considered to be due to
solar polar magnetic activity, the simulation of long term
solar activity due to poloidal oscillations in Fig 1 shows also normal activity
during deep minimum period activity confirming
the observations.  
\section {Conclusions}
In this talk, after summarizing the solar observations,
two theoretical models, viz., {\em turbulent dynamo} and
{\em MHD oscillations} mechanisms on the genesis of
solar cycle and activity phenomena are critically examined. The seminal
work of Alfven on the solar cycle is revisited. The new ideas
on the genesis of the solar cycle and activity phenomena
and it's long-term variations are presented. 

 \begin{center}
{\Large\bf Acknowledgments}
\end{center}
Author is thankful to Dr. Katya Georgieva, Dr. Haubold and SOC for
providing full financial support for attending the conference.
Author is also thankful to Prof. Stenflo for the useful 
discussions.

\medskip
\section {\Large\bf References}
\noindent Alfven H 1943 {\it Arkiv Math Astr Fys} {\bf 29A} No 12

\noindent Altrock R Howe R and Ulrich R 2008 {\it ASP Conf Ser} {\bf 383} p. 335

\noindent P. Ambroz, M. Druckmüller, A. A. Galal and R. H. Hamid, 2009, Sol. Phys., 258, 243

\noindent Antia, H. M. \& Basu, S., 2000, ApJ, 541, 442

\noindent Antia, H. M., Chitre, S. M., \& Thompson, M. J. 2000, A\&A, 360, 335

\noindent Antia, H. M. 2002, Proceedings of IAU Coll. 188, ESA SP-505, p.71

\noindent Antia, H. M., Basu, S. and Chitre, S. M., 2008 {\it ApJ} 681 680

\noindent Babcock H W 1961 {\it ApJ} {\bf 133} 572

\noindent Basu, S. 1997, {\it MNRAS}, 288, 572

\noindent Beck, C. and Rezaei, R., 2009, A\&A, {\bf 502}, 969

\noindent Beer J Tobias S Weiss W 1998 {\it Sol Phys} {\bf 181} 237

\noindent Braithwaite J and Spruit H C {\it Nature} {\bf 431} 819

\noindent Brandenburg A 1988 in {\it Proc of 6th Sov-Fin Astron Meet} p. 34 

\noindent Brandenburg A and Subramaniyam K 2005 {\it Phys Rep} {\bf 417} p.1 

\noindent Brandenburg A and Speigel I A 2008 {\it Astron Nach} {\bf 329} 351

\noindent Brun {\em et. al.}, ApJ, 614, 1073, 2004

\noindent Bushby P J Houghton S M Proctor M R E and Weiss N O 2008 MNRAS 387 698

\noindent Charbabonneau P and Dikpati M 2000 {\it ApJ} {\bf 543} 1027 

\noindent Charbabonneau P 2005 {\it Living Rev Sol Phys} {\bf 2}

\noindent Choudhuri, A. R., 2008, Advances in Space Research, 41, 868

\noindent Cliver E W Boriakoff V and Feyman J 1998 {\it Geophys Res Let} {\bf 25} 1035  

\noindent Choudhuri A R 1992 {\it A\&A} {\bf 253} 277

\noindent Choudhuri A R Chatterjee P Jiang J 2007 {\it Phys Rev Let} {\bf 98} 131103

\noindent Choudhuri A R 2008 {\it Adv in Space Res} {\bf 41} 868

\noindent Christensen-Dalsgaard, J., 2002, Int. Journ. Mod. Phys. D, 11, 995-1009

\noindent Christensen-Dalsgaard, J., 2003, Lecture Notes on "Stellar Oscillations"

\noindent Cowling T G 1934 {\it MNRAS} 94 39

\noindent Cowling T G 1953 in {\it The Sun} p. 532

\noindent Cowling T G  1981 {\it Ann Rev Astron Astrophys} {\bf 19} 115

\noindent De Jager C and Versteegh G M 2005 {\it Sol Phys} {\bf 229} 175

\noindent Davila J M and Chitre S M 1996 {\it Bull Astron Soc India} {\bf 24} 309

\noindent De Jager C 2005 {\it Space Sci Rev} {\bf 120} 197

\noindent De Luca E E and Gilman P A  1991 in {\it Solar Interior Atmosphere} p. 275

\noindent Dicke R H 1978 {\it Nature} {\bf 276} 676

\noindent Dikpati M 2005 {\it Advances in Space Res}, {\bf 35}, p.322 	

\noindent Dikpati M and Gilman P A 2006 {\it ApJ}, {\bf 649}, 498 	

\noindent Feynman J 1982 {\it JGR} {\bf 87} 6153

\noindent Feynman J 1983 {\it Rev of Geophys and Space Phys} {\bf 21} 338

\noindent Fletcher N H and Rossing T D 1998 {\it The Physics of 
Musical Instruments}, second edition, p. 103 

\noindent Garcia, R. A., {\em et. al.}, 2008; {\it Astronom Nach}, {\bf 329}, 476

\noindent Georgieva K and Kirov B 2006 {\it Sun and Geosphere} {\bf vol 1}, 12 

\noindent Gilman P A and Miesch M S 2004 {\it ApJ} {\bf 611} 568

\noindent Gokhale M H Javaraiah J Hiremath K M 1990 {\it IAU Symp} {\bf 138} p. 375

\noindent Gokhale M H and Javaraiah J 1992 {\it Sol Phys} {\bf 138} p. 399

\noindent Gough D O and McIntyre M E 1998 {\it Nature} {\bf 394} 755

\noindent Grandpierre A and G'bor G 2005 {\it Astrophys\& Space Sci} {\bf 298} 537

\noindent Hale G E 1908 {\it ApJ} {\bf 28} 315

\noindent Hasan, S. S., 2008, in {\em Physics of the Sun and its Atmosphere}, 
eds. B.N. Dwivedi and U. Narain, p. 9

\noindent Hathaway D H Nandy D Wilson R M and Reichmann E J 2004 {\it ApJ} {\bf 602} 543

\noindent Hiremath K M 1994 {\it Ph D Thesis} Bangalore University, India 

\noindent Hiremath K M and Gokhale 1995 {\it ApJ} {\bf 448} 437

\noindent Hiremath K M 2001 {\it Bull Astron Soc India} {\bf 29} 169

\noindent Hiremath K M 2006 {\it A\&A} {\bf 452} 591 

\noindent Hiremath K M and Lovely M R 2007 {\it ApJ} {\bf 667} 585

\noindent Hiremath K M 2008a {\it Astrophyhs \& Space Sci} {\bf 314} 45

\noindent Hiremath K M 2008b {\it arXiv:0803.1242} eprint

\noindent Hiremath K M 2009a, accepted in Sun and Geosphere, also see the eprint {\it arXiv:0906.3110}

\noindent Hiremath K M 2009b, {\it arXiv:0909.4376} eprint

\noindent Javaraiah J and Gokhale M H 1995 {\it Sol Phys} {\bf 158} 173

\noindent Howe, R., Cristensen-Dalsgaard, J., Hill, F., {\em et. al.,} 2000, Science, 287, 2456

\noindent Howe, R., 2009, Living Reviews in Solar Physics, vol. 6, no. 1

\noindent Javaraiah, J and Komm, R. W., 1999, Sol. Phys, 184, 41

\noindent Javaraiah, J., 2000, Ph. D. Thesis, Study of Sun's rotation and 
solar activity, Bangalore University, India

\noindent Jimenez, A. and Garcia, R. A., 2009, Accepted for publication in ApJSS

\noindent Jin, Chunlan; Wang, Jingxiu and Zhou, Guiping, 2009, {\it ApJ}, 697, 693

\noindent Jouve and Brun, A\&A, 474, 239, 2007
                                                                                                              
\noindent Howard R and La Bonte B J 1980 {\it ApJ} 239 L33

\noindent Knaack R and Stenflo J O 2005 {\it A\&A} {\bf 438} 349

\noindent Komm R W Howard R F Harvey J W 1993 {\it Sol Phys} {\bf 143} 19

\noindent Krivova, N. A. \& Solanki, S. K., 2002, A \& A, 394, 70

\noindent Layzer D Krook M and Menzel D H 1955 {\it Proc Roy Soc} {\bf A223} 302

\noindent Layzer D Rosner R and Doyle H T  1979 {\it ApJ} {\bf 229} 1126

\noindent Legrand J P and Simon P A 1991 {\it Sol Phys} {\bf 131} 187

\noindent Leighton R B 1964 {\it ApJ} {\bf 140} 1547

\noindent Leighton R B 1969 {\it ApJ} {\bf 156} 1

\noindent Levy E H 2002 in {\it The Solar Cycle} ASP Conf Ser {\bf 27} p. 139

\noindent K. J. Li, Q. X. Li, T. W. Su, and P. X. Gao, 2006,
Sol Phys, {\bf 239}, 493

\noindent Lites, B. W.; Leka, K. D.; Skumanich, A.; Martinez Pillet, V and 
Shimizu, T, 1996; {\it ApJ}, {\bf 460}, 1019 

\noindent Lites, B. W., {\em et. al.}, 2008, {\it ApJ}, {\bf 672}, 1237

\noindent Lites, B. W., {\em et. al.}, 2009, First Results From Hinode ASP Conference Series, {\bf 397},
p.17

\noindent Mestel L and Weiss N O 1987 {\it MNRAS} {\bf 226} 123

\noindent Mark S. Miesch and Juri Toomre, Annual Review of Fluid Mechanics,  
Vol. 41: 317-345, 2009

\noindent Moss D Sokoloff D Usoskin I and Tutubalin V 2008 {\it Sol Phys} {\bf 250} 221

\noindent Muscheler R et. al. 2007 {\it Queternary Science Reviews} {\bf 26} 82

\noindent Nandy D and Choudhuri A R 2002 {\it Science} {\bf 296} 167

\noindent Nandy, D., arXiv:0906.4748, 2009

\noindent Obridko, V. N and Shelting, B. D., 2007, Advan in Space Res, 40, 1006

\noindent Pasachoff, J. M., 2009, {\it Nature} {\bf 459}, 789

\noindent J. M. Pasachoff1,, V. Rušin, M. Druckmüller, P. Aniol, M. Saniga and M. Minarovjech,
2009, {\em ApJ}, 702, 1297 

\noindent Parker E N 1955 {\it ApJ} {\bf 122} 293

\noindent Pelt J Brooks J Pulkkinen P J Tuominen I 2000 {\it A\&A} {\bf 362} 1143

\noindent Petrovay K 2000 {\it ESA SP} {\bf 463} p. 3

\noindent Piddington J H 1971 {\it Proc Astron Soc Australia} {\bf 2} 7

\noindent Piddington J H 1972 {\it Sol Phys} {\bf 22} 3

\noindent Plumpton C and Ferraro V C A 1955 {\it ApJ} 121 168

\noindent Piddington J H 1973 {\it Astrophys Space Sci} {\bf 24} 259

\noindent Price C P Prichard D and Hogenson E A 1992 {\it JGR} {\bf 97} 19113

\noindent Priest E R 1981 in {\it Solar magneto-hydrodynamics}

\noindent Shirley J M 2006 {\it MNRAS} {\bf 368} 280

\noindent Solanki S K Inhester B and Schussler M 2006 {\it Rep Prog Phys} {\bf 69} 563

\noindent Spruit H C 1990 in {\it Inside the sun} 415

\noindent Stein R F and Nordlund A 2006 {\it ApJ} 642 1246

\noindent Steiner O Rezaei R Schaffenberger W and Wedenmeyer-Bohn S 2008 {\it ApJ} 680 85  

\noindent Stenflo J O and Vogel M 1986 {\it Nature} {\bf 319} 285

\noindent Stenflo J O 1988 {\it Astrophys Space Sci} {\bf 144} 321

\noindent Stenflo J O and Gudel M 1988 {\it A\&A} {\bf 191} 137

\noindent Stenflo J O 1994 in {\it Solar surface magnetism} p. 365

\noindent Svanda M Kosovichev A G and Zhao J {\it ApJ} {\bf 670} 69

\noindent Toomre, J., Christensen-Dalsgaard, J., Hill, F., Howe, R., Komm, R. W.,
 Schou, J. and Thompson, M. J., 2003, in "Proceedings of SOHO 12 / GONG+ 2002", 
ESA SP-517, p. 409

\noindent Tsuneta, S., {\em et. al.}, 2009, {\it ApJ} {\bf 688}, 1374

\noindent Unno, W., Osaki, Y., Ando, H and Shibahashi, H, 1979, in "Nonradial oscillations of stars",
University of Tokyo Press

\noindent Usoskin I G Solanki S K and Kovaltsov G A {\it A\&A} {\bf 471} 301

\noindent Vainstein S I and Cattaneo F 1992 {\it ApJ} {\bf 393} 165

\noindent Vandakurov Y V 1990 {\it IAU Symp} {\bf 138} 333

\noindent Vecchio, A and Carbone, V, A\&A, 2009, {\bf 502}, 981

\noindent Venkatakrishnan, P and Gosain, S, 2008, in {\em Physics of the Sun and its Atmosphere},
eds. B.N. Dwivedi and U. Narain, p. 39

\noindent Walen C 1949 in {\it On the Vibratory Rotation of the Sun} 

\noindent Wijn,D. A. G.; Stenflo, J. O.; Solanki, S. K. and Tsuneta, S., 2009, {\it Space Science Rev}, 
{\bf 144}, 275

\noindent Wilson I R G Carter B D and Waite I A 2008 {\it Pub Astron Soc Austr} {\bf 25} 85

\noindent Zaqarashvili T V 1997 {\it ApJ} {\bf 487} 930

\noindent Zhao J and Kosovichev A G 2004 {\it ApJ} {\bf 603} 776

\noindent Zwaan C 1981 in {\it The sun as a star} {\bf NASA SP-450} p.163
 
\end{document}